\documentclass[conference]{IEEEtran}
\usepackage{amsmath}
\usepackage{amssymb}
\usepackage{color,soul}
\usepackage[pdftex]{graphicx}
\allowdisplaybreaks[4]
\usepackage[noadjust]{cite}

\usepackage{float}
\usepackage{comment}
\usepackage{epstopdf}
\usepackage{pgfplots, pgfplotstable}
\pgfplotsset{compat=newest}
\usetikzlibrary{plotmarks}
\usetikzlibrary{arrows.meta}
\usetikzlibrary{positioning,shapes,shadows,arrows,shapes.multipart}
\usepgfplotslibrary{fillbetween}
\usepgfplotslibrary{patchplots}
\usepgfplotslibrary{groupplots}
\usepackage{grffile}
\usepackage{url}
\usepackage[nolist,printonlyused]{acronym}
\usepackage{listings}
\newfloat{lstfloat}{t}{lop}
\floatname{lstfloat}{Code Block}

\usepackage{algorithm}
\usepackage{algorithmic}
\usepackage{url}
\usepackage{xfrac}
\usepackage{bbm}
\newcommand{\scale}{0.94}
\usepackage[font=small,skip=3pt]{caption}
\renewcommand{\baselinestretch}{0.98}
\newtheorem{remark}{Remark}

\begin{document}

\title{Deep Reinforcement Learning for DER Cyber-Attack Mitigation}

\author{\IEEEauthorblockN{Ciaran Roberts \\ Sy-Toan Ngo, Alexandre Milesi \\ Sean Peisert, Daniel Arnold}
\IEEEauthorblockA{\textit{Lawrence Berkeley National Laboratory}\\
\{cmroberts,sytoanngo,amilesi,\\sppeisert,dbarnold\}@lbl.gov}
\and
\IEEEauthorblockN{Shammya Saha\\ Anna Scaglione \\ Nathan Johnson}
\IEEEauthorblockA{\textit{Arizona State University}\\
\{sssaha,ascaglio, \\ nathanjohnson\}@asu.edu}
\and
\IEEEauthorblockN{Anton Kocheturov\\ Dmitriy Fradkin}
\IEEEauthorblockA{\textit{Siemens Corporation} \\ \textit{Corporate Technology}\\
\{anton.kocheturov,\\dmitriy.fradkin\}@siemens.com}
}
\maketitle
\let\thefootnote\relax\footnote{This research was supported in part by the Director, Cybersecurity, Energy Security, and Emergency Response, Cybersecurity for Energy Delivery Systems program, of the U.S. Department of Energy, under contract DE-AC02-05CH11231.  Any opinions, findings, conclusions, or recommendations expressed in this material are those of the authors and do not necessarily reflect those of the sponsors of this work.}

\vspace{-0.3in}

\begin{abstract}
The increasing penetration of DER with smart-inverter functionality is set to transform the electrical distribution network from a passive system, with fixed injection/consumption, to an active network with hundreds of distributed controllers dynamically modulating their operating setpoints as a function of system conditions. This transition is being achieved through standardization of functionality through grid codes and/or international standards. DER, however, are unique in that they are typically neither owned nor operated by distribution utilities and, therefore, represent a new emerging attack vector for cyber-physical attacks. Within this work we consider deep reinforcement learning as a tool to learn the optimal parameters for the control logic of a set of uncompromised DER units to actively mitigate the effects of a cyber-attack on a subset of network DER.
\end{abstract}


\section{Introduction}
\label{sec:intro}

The increasing penetration of distributed energy resources (DER) in electrical distribution systems is causing a paradigm shift in how these networks are managed. While these systems were historically passive, distributed power generation is forcing distribution grids to become more dynamic as DER are  expected to provide grid services, \textit{e.g.} voltage control. This transition presents several challenges, particularly in the area of cyber-physical security \cite{qi2016cybersecurity, sahoo2019cyber}.

DER are especially unique when it comes to cyber-physical security. These devices are typically neither utility owned nor directly controlled and, therefore, present a new attack vector for adversaries seeking to disrupt normal grid operating conditions. Additionally, many manufacturers and/or aggregators remotely control large populations of these devices via cellular networks, customers' WiFi routers, or wired internet connections \cite{hawai}. This makes ensuring the integrity of commands significantly more difficult. While recent standards (e.g. IEEE 1547 standard) seek to specify minimal control requirements for these devices, they do not explicitly address the associated cyber-physical security challenges \cite{epri1547}. Inverter manufacturers and aggregators have the ability to remotely monitor and control the settings for inverters/DER deployed in the field. Once access to the central system is gained, that system can be used to push malicious control logic back to all DER.  Thus, utilities have already expressed concerns about how the impact of a single cyber intrusion into an aggregators'/manufacturers' internal network could be exploited to compromise an aggregator’s/manufacturer’s entire DER fleet \cite{hawai}. In regions with high penetration of these devices, this could have devastating effects. 

In this work we adopt a purely physics-based approach for the mitigation of cyber-physical attacks on DER (specifically, solar photovoltaic inverters). We assume that the adversary has already gained access to a subset of DER on a given network and seeks to maliciously re-configure the control settings of smart inverters to disrupt distribution grid operations. Our approach does not focus on detecting the cyber-intrusion but rather mitigating the resulting physical manifestation of the attack on the grid. To develop optimal control policies that mitigate the impact of these attacks, we train a deep reinforcement learning (DRL) policy that re-configure the control settings of uncompromised DER. This trained policy is then deployed locally on controllable DER and determines smart inverter parameter updates based on locally observed information.

DRL has been gaining increasing attention in recent years, including in power systems, for determining control policies for highly complex non-linear systems. In \cite{huang2019adaptive}, the authors use  Deep Q-Network (DQN) learning, a reinforcement learning (RL) algorithm that combines Q-Learning with deep neural networks, to control both generator dynamic braking and load shedding in the event of a contingency to ensure post-fault recovery.  In \cite{yang2019twotimescale}, the authors consider the problem of coordinated voltage regulation using capacitors and smart inverters. Exploiting the timescale separation of these devices, they solve a convex optimization problem to determine the control policies of the smart inverters while using a DQN network to learn an optimal policy for capacitor bank switching. In \cite{li2019coordination}, a deep deterministic policy gradient (DDPG) RL agent is used to co-ordinate across DER and directly modulate active and reactive power to regulate the grid voltage during normal operations.

This work differs from those described above in that we focus on developing a supervisory control policy that continuously monitors system conditions and takes action during sustained abnormal behavior. This controller, therefore, should not impact an inverters response to normal disturbances, e.g. line-to-ground faults. While the proposed controller design is motivated by the need to respond to cyber-physical attacks, it is  agnostic to the cause of the abnormal conditions. Consequently, it can also serve to autonomously re-configure controller settings in the event that an intended action has resulted in abnormalities, for instance, when connecting different microgrids with independently optimized controllers.  This paper presents a framework for DRL for smart grid applications and explores the use case of a cyber-physical attack intended to induce oscillatory behavior in the grid voltage.  

The remainder of the paper is organized as follows. Section~\ref{sec:drl} gives a brief introduction to DRL and the terms that will be used throughout the paper. Section~\ref{sec:models} gives an overview of the power system models and networks used in the study. Finally, Section~\ref{sec:results} presents the results and Section~\ref{sec:conclusions} summarizes some of the key conclusions.

\section{Deep Reinforcement Learning}
\label{sec:drl}
\subsection{Reinforcement Learning}

\begin{figure}[H]
    \centering
    \resizebox {18pc} {!} { \begin{tikzpicture}
\tikzstyle{mynode}=[draw, align=center, rounded corners=5pt,minimum width=3cm, minimum height=1cm]
\node[mynode] (env) at (0,0) {Environment};
\node[mynode] (agent) [above=of env] {Agent};

\draw[bend left=90, -latex, align=center] (env.west) to node [auto] {State: ($s_t \in \mathcal{S}) \sim \mathcal{P}$\\ Reward: $R_t \in \mathcal{R}$} (agent.west);
\draw[bend left=90, -latex, align=center] (agent.east) to node [auto] {Action \\($a_t \in \mathcal{A}) \sim \pi$} (env.east);
\end{tikzpicture} }
    \caption{Reinforcement learning loop.}
    \label{fig:rl_loop}
\end{figure}
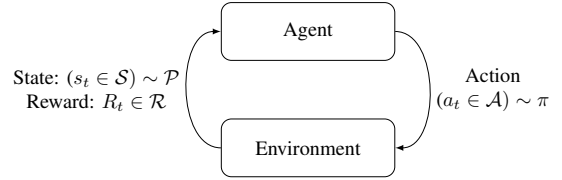

RL is a branch of machine learning focused on optimal decision making in stochastic environments.  The goal of RL techniques is to train an agent (\textit{i.e.} the decision-maker) to interact with an environment in such a way as to maximize a cumulative reward.  The environment is usually cast as a Markov Decision Process (MDP), which consists of the following elements:
\begin{itemize}
    \item A state space, $\mathcal{S}$, containing states observable by the agent;
    \item An action space, $\mathcal{A}$, containing all the possible actions the agent can execute;
    \item A state transition function, $\mathcal{P}: \mathcal{S} \times \mathcal{A} \times \mathcal{S} \rightarrow [0, 1]$, specifying the probability distribution over the next state $s'$ when an action $a$ is taken at state $s$;
    \item A reward function, $\mathcal{R}: \mathcal{A} \times \mathcal{S} \times \mathcal{S} \rightarrow \mathbb{R}$, specifying the reward received by the agent when the environment transitions from state $s$ to state $s'$ with action $a$; 
    \item A discount factor, $\gamma \in [0, 1]$, representing the trade-off between immediate and future rewards.
\end{itemize}
A RL agent learns optimal actions by repeatedly interacting with the environment and assessing the value of resulting rewards, $R_t \in \mathbb{R}$, dependent on the actions taken, $a_t \in \mathcal{A}$, and the states of the environment, $s_t \in \mathcal{S}$. The agent-environment interaction is visualized in Fig. \ref{fig:rl_loop}.  As shown in the figure, the agent takes action $a_{t}$ following policy $\pi$ causing a state transition in the environment.  The new state, $s_{t}$, and subsequent reward, $R_{t}$, are observed by the agent and can then be used to update the policy $\pi$. The  objective of the agent is to maximize the discounted reward $J(\pi) = \mathbb{E}_\pi \left[ \sum_{t=0}^{T} \gamma^t R_t\right]$, where $T$ is the terminal time step, by following a policy $\pi$ which can be deterministic or stochastic in nature.

\subsection{Deep Reinforcement Learning}

Classical RL relies on feature engineering and is difficult to apply to environments with high dimensional, continuous action and/or state spaces \cite{sutton2018reinforcement}.  Such spaces, typically, must be discretized first, leading to a combinatorial explosion in complexity and unreasonable training time (the so-called \textit{curse of dimensionality}). In addition, classical RL has trouble capturing patterns in the presence of noisy or incomplete data. DRL solves these issues by leveraging neural networks with multiple hidden layers that take the agent observations as input and output a  policy that determines what action to take in a given state.

With DRL, the inputs to the neural network can be structured data (tabular data), unstructured data (images, text, video), or both. The weights of these neural networks are efficiently learned end-to-end via gradient-based optimization to find the best intermediate features and an optimal output policy. The need for precise feature engineering is then greatly reduced, thanks to the automatic high-dimensional feature extraction of the hidden layers. In DRL, one can use the networks to explicitly approximate an optimal policy distribution, $\pi$, over possible actions. This distribution is then sampled by the agent to determine the next action, as in policy gradient methods. They may also be used to approximate either a value function, $V^\pi(s)$, or an action-value function, $Q^\pi(s,a)$, from gathered data, leading to an action decision based on inferred values for all possible future states, as in DQN. The value function, $V^\pi(s)$, is the expected discounted reward when starting in state $s$ and following the policy $\pi$, whereas the action-value function $Q^\pi(s,a)$ is defined as the expected discounted reward when starting in state $s$, taking action $a$, and then following the policy $\pi$ thereafter.

Thanks to its flexibility, DRL has been successfully applied to robotic control \cite{openai2018learning}, video games \cite{mnih2013playing, openai2019dota} and board game playing  \cite{silver2016mastering, silver2018general}.

\subsection{Policy Gradient and PPO} \label{sec:ppo}
Policy gradient methods employ a policy modeled by a neural network which is trained directly by gradient ascent on the expected return. The most basic method (vanilla policy gradient) is simple to implement but has the drawback of having a high gradient variance. In response, Actor-Critic (AC) methods were proposed\cite{mnih2016asynchronous}, where another, possibly shared, neural network approximates the value function.

Let $\pi_{\theta}(a|s)$ be a stochastic policy, parameterized by $\theta$, modeling the probability distribution of action $a \in \mathcal{A}$ given the state $s \in \mathcal{S}$. Let $V^{\pi}_{\phi}(s)$ be a value function parameterized by $\phi$, estimating the cumulative discounted reward from the current state to the terminal state. The gradient of $J(\theta)$ is:
\begin{equation}
\nabla_\theta J(\theta) = \underset{{\tau \sim \pi_\theta}}{\mathbb{E}}\!\left[\sum_{t=0}^{T}\nabla_\theta \log\pi_\theta(a_t|s_t)A_{\phi}^{\pi}(s_t, a_t)\right],
\end{equation}
where $\tau$ is the trajectory generated by policy $\pi_\theta$ and $A_{\phi}^{\pi}(s_t, a_t) = R_t + \gamma V_{\phi}^\pi(s_{t+1})- V_{\phi}^\pi(s_t)$ is the advantage estimation, representing how much better taking action $a_t$ is, as opposed to following the policy $\pi$ when in state $s_t$. The policy and value function are updated by gradient ascent/descent:
\begin{align}
&\quad \theta_{k+1} = \theta_{k} + \alpha \nabla_{\theta} J(\theta),\\
\phi_{k+1} = &\phi_{k} - \beta \nabla_{\phi}(R_t + V_\phi^\pi(s_{t+1}) - V_\phi^\pi(s_t))^2.
\end{align}

As the training of AC methods can be unstable when the data distribution changes due to a large policy update, the Trust Region Policy Optimization (TRPO) was introduced \cite{schulman2015trust}. TRPO limits the updates in the policy space by enforcing a Kullback–Leibler divergence constraint on the size of each update. A Proximal Policy Optimization (PPO) \cite{schulman2017proximal}  using a clipped surrogate objective simplifies the aforementioned method and yields similar performance:
\[
\begin{gathered}
L^{\text{CLIP}}(\theta)=\hat{\mathbb{E}}\left[  \min \Big(r_t(\theta)\hat{A}_t, \  \text{clip}\big(r_t(\theta), 1-\epsilon, 1+\epsilon\big)\hat{A}_t\Big)\right], \\
\text{where}\  r_t(\theta) \triangleq \frac{\pi_\theta(a_t | s_t)}{{\pi_\theta}_{\text{old}}(a_t | s_t)} \ \text{and}\ \hat{A}_t\triangleq A_{\phi}^{\pi}(s_t, a_t)
\end{gathered}
\]
This clip operation encourages a more gradual updates to the policy rather than large changes, and the minimum between the unclipped and the clipped objective is used so that the final objective is a lower bound on the unclipped objective \cite{schulman2017proximal}. The hat over the expectation means that we compute a Monte Carlo estimate of it.

PPO is a state-of-the-art method that was successfully used in video games \cite{openai2019dota} and robotics in simulation \cite{heess2017emergence}. We consider here its application to the control of smart inverters.
Before we map the specific problem onto the RL formalism, the following remark is in order:
\begin{remark}
In many applications, the state of the entire system, $s_t$, is not directly observed.
In this case, the problem falls in the class of Partially Observable MDPs (POMDP). In a POMDP, the additional element in the model is:
\begin{itemize}
    \item An observation  transition function (also called perceptual distribution or emission probability) ${\cal V}~:~ {\cal S}\times {\cal O}\rightarrow [0,1]$ that specifies the probability distribution of the observation $o_t$ given the state $s_t$.
\end{itemize}
The policy function in this case takes as input the observation rather than the state, i.e. the goal is to find the optimum $\pi(a_t|o_t)$. As mentioned later, the formulation in this paper falls in the class of POMDP. 
Also, we note that neither the state transition function nor the perceptual distribution are explicitly given; hence the policy neural network is trained through a Monte Carlo method.  
\end{remark}

\section{Methodology}
\label{sec:models}

\subsection{Modeling the DER action space}
\begin{figure}[h!]
\centering
\resizebox {17pc} {!} { \begin{tikzpicture}

\coordinate (xmax) at (5,0);
\coordinate (xmin) at (-5,0);
\draw[<->] (xmin) -- (xmax) node[anchor=west] {$V$};
    
\coordinate (ymax) at (0,3);
\coordinate (ymin) at (0,-3);
\draw[<->] (ymin) -- (ymax) node[anchor=south] {\% available VARs};
    
\filldraw[black]	
		(-2,0) circle (2pt) node[anchor=north] {$\eta_{2}$}
		(2,0) circle (2pt) node[anchor=south] {$\eta_{3}$}
        
        (-4,2) circle (2pt)
        (4,-2) circle (2pt);
        
\draw	
		(0,2) node[anchor=west] {100\%}
		(-4,0) node[anchor=north] {$\eta_{1}$}
		(0,-2) node[anchor=east] {-100\%}
		(4,0) node[anchor=south] {$\eta_{4}$};

\draw[dotted] (4,0) -- (4,-2);
\draw[dotted] (-4,0) -- (-4,2);

\draw[dotted] (0,-2) -- (4,-2);
\draw[dotted] (0,2) -- (-4,2);

\draw[thick][<->] (-5,2) -- (-4,2) -- (-2,0) -- (2,0) -- (4,-2) -- (5,-2);

\end{tikzpicture} }
\caption{Inverter Volt-VAR curve.  Positive percent of VAR injection.}
\label{fig:vvc}
\end{figure}
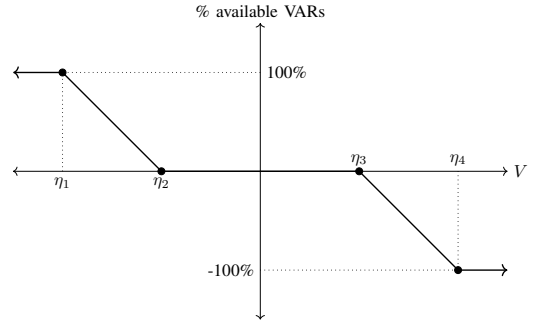

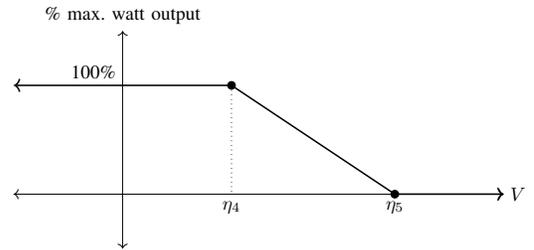
\begin{figure}[h!]
\centering
\resizebox {17pc} {!} { \begin{tikzpicture}

\coordinate (xmax2) at (7,0);
\coordinate (xmin2) at (-2,0);
\draw[<->] (xmin2) -- (xmax2) node[anchor=west] {$V$};
    
\coordinate (ymax2) at (0,3);
\coordinate (ymin2) at (0,-1);
\draw[<->] (ymin2) -- (ymax2) node[anchor=south] {\% max. watt output};
    
\draw
        (2,0) node[anchor=north] {$\eta_{4}$}
        (0,2.25) node[anchor=east] {100\%};

\filldraw[black]
        (5,0) circle (2pt) node[anchor=north] {$\eta_{5}$}
        (2,2) circle (2pt);

\draw[dotted] (2,2) -- (2,0);

\draw[thick][<->] (-2,2) -- (2,2) -- (5,0) -- (7,0);

\end{tikzpicture} }
\caption{Inverter Volt-Watt curve.  Positive percent of watt injection.}
\label{fig:vwc}
\end{figure}

In response to evolving standards and requirements, DER are increasingly being deployed with the ability to modulate their real and reactive power injection/consumption in response to locally measured grid conditions. In this work we focus specifically on smart inverter Volt-VAR (VV) and Volt-Watt (VW) control functionality as these operating modes are designed to help regulate distribution system voltages in the presence of large amounts of renewable generation. Under VV/VW control schemes, each inverter seeks to modulate active and reactive power injections in response to measured system voltage.  The amount by which reactive and active power injections are modulated is governed according to piece-wise linear functions of voltage, often referred to as ``droop" curves.
Different parameterizations of VV and VW curves exist, however, existing guidelines often depict shapes shown in Figs. \ref{fig:vvc} - \ref{fig:vwc}, which are parameterized by the five parameters that define the piece-wise linear curves shown, which will be referred to as the components of the setpoint-vector $\eta=[\eta_{1}, \ldots,\eta_{5}]$. In this work, the action is a $5\times 1$ vector $a=\Delta\eta\triangleq\eta-\eta^o$, where $\eta^o$ is the default set of parameters. Note that, even though in principle the action is continuous, we quantize the possible range for the action and search directly for the categorical vector $a$.

The VV curve injects reactive power when voltages in the system are low and transitions to VAR consumption as voltages increase.  The VW curve provides maximum real power injection under most voltage levels, but curtails PV output as voltage levels increase.  The additional capacity resulting from active power curtailment can then be used for additional reactive power consumption.

Without loss of generality, we assume all inverters in the subsequent analysis possess both VV and VW functionality.  
Let $p^{\text{max}}$ be the maximum output of the PV unit under presently available solar insolation, and $q^{\text{avail}}$ the limit for reactive power in absolute value.  
In some instances, the amount of reactive power available for injection/consumption may be fixed (in the case of an oversized inverter relative to the capacity of the PV panels) while in others, $q^{\text{avail}}$ may depend on the amount of real power being generated from the PV system:
\begin{align}
    q^{\text{avail}} \le \sqrt{ s^{2} - f^{p}(\bar v)^{2} } \label{eq:q_avail},
\end{align}
where $s$ is the inverter capacity.  
Let $u_{p,i}$ and $u_{q,i}$ denote the active and reactive power control signal of inverter $i$. They are function of the {\it averaged} measured voltage magnitude at the bus (c.f. \eqref{eq:vmdot}). Rather than considering completely arbitrary VV and VW mappings $u_{p,i}$ and $u_{q,i}$ that respect the limits $p^{\max}$ and $q^{\mbox{avail}}$, we seek policies that are expressed as:
\begin{align}
    u^{p}_{i} &= f^{p}_{i}(\bar v)\triangleq
        \begin{cases}
                    p^{\text{max}} & \bar v \in [0, \eta_{4}] \\[0.5em]
                    \left( \frac{ \eta_{5} -  \bar{v}}{ \eta_{5} -  \eta_{4}} \right) p^{\text{max}} &  \bar{v} \in ( \eta_{4},  \eta_{5}]
                    \\[0.5em]
                    0 & \bar v \in ( \eta_{5},\infty)
                \end{cases}. \label{eq:p}
\end{align}
\begin{align}
    u^{q}_{i} &= f^{q}_{i}(\bar v)\triangleq
        \begin{cases}
        q^{\text{avail}} & \bar v \in [ 0,\eta_{1}]\\[0.5em]
        \left( \frac{ \eta_{2} - \bar v }{ \eta_{2} -  \eta_{1}} \right) q^{\text{avail}} & \bar v \in ( \eta_{1},  \eta_{2}] \\[0.5em]
        0 &  \bar v \in (\eta_{2},  \eta_{3})\\[0.5em]
        -\left( \frac{ \eta_{3} - \overline{v} }{ \eta_{4} -  \eta_{3}} \right) q^{\text{avail}} & \bar v \in [ \eta_{3},  \eta_{4}] \\[0.5em]
        -q^{\text{avail}} & \bar v \in ( \eta_{4},\infty)
    \end{cases}, \label{eq:q}
\end{align}

The scheme of \eqref{eq:p} - \eqref{eq:q_avail} illustrates the combined use of VV and VW control with VW  precedence \cite{inverter2016}.  Under VW precedence, priority is given to the VW controller to determine any needed curtailment before determining the VARs available ($q^{\text{avail}}$).  After $q^{\text{avail}}$ is fixed, $ u^{q}_{i}$ is computed from \eqref{eq:q}. 

In the event of a cyber-physical attack we assume that an adversary has the capability to re-dispatch a set of voltage breakpoints $\eta=[\eta_{1}$, \ldots $\eta_{5}]$ that parametrize the droop curves in Figs. \ref{fig:vvc} - \ref{fig:vwc} for a subset of DER on the network. Within the context of this work, the remaining set of non-compromised DER can then be updated with new parameters vector $\eta'=a+\eta^o$ to re-shape their own local droop curves to transition the system voltages to a \textit{safe} region, devoid of oscillatory behavior.


Finally, the structure of the DER VV and VW control dynamic response, similarly to  \cite{farivar2013equilibrium, braslavsky2017voltage, inverter2016},  includes the following first order low pass filters that average the input voltage and determine the active and reactive power injections:
\begin{subequations}
\begin{align}
    \bar{v}_{i,t} &= \bar{v}_{i,t-1} + \tau_{i}^{m}\big(v_{i,t} - \bar{v}_{i,t-1}\big) \label{eq:vmdot},\\
     p_{i,t} &= p_{i,t-1} + \tau_{i}^{o}\big(f^{p}_{i}(\bar{v}_{i,t}) - p_{i,t-1}\big), \label{eq:p_q} \\
     q_{i,t} &= q_{i,t-1} + \tau_{i}^{o}\big(f^{q}_{i}(\bar{v}_{i,t}) - q_{i,t-1}\big) \label{eq:q_g},
\end{align}
\end{subequations}


\noindent where $\bar{v}_{i}$ denotes a low-pass filtered measured of the voltage magnitude, $v_{i}$, at node $i$,  $\tau^{m}_{i}$ is its associated measurement time constant, $\tau^{o}_{i}$ is the output filter time constant and $f^{p}_{i}(\bar{v}_{i,t})$ and $f^{q}_{i}(\bar{v}_{i,t})$ are the piecewise linear functions of the measured nodal voltages for node $i$ given by \eqref{eq:p} and \eqref{eq:q} respectively. 
Note the equilibrium of \eqref{eq:p_q} - \eqref{eq:q_g} is given by \eqref{eq:p} - \eqref{eq:q}.  

The stability of \eqref{eq:vmdot} - \eqref{eq:q_g} has been studied in \cite{farivar2013equilibrium} and \cite{bakerNetwork2017}, where it has been observed that instabilities manifest as oscillations in inverter power injections and nodal voltages.  

As said before, the RL agent indirectly manipulates the outputs of the inverter by modifying the vector of parameters  $\eta_t=a_t+\eta^o$.  
Next we define a component of the observation $o_t$ used as an input to the DRL controller in our POMDP formulation. The quantity is a local measure of the presence and severity of voltage magnitude aforementioned oscillations. %

\subsection{A Measure of Unstable Oscillations} \label{subsec:detect}
\begin{figure}[ht]
\centering
\resizebox {21pc} {!} { \tikzstyle{block} = [draw, thick, rectangle, 
    minimum height=3.5em, minimum width=4em]
\tikzstyle{triang} = [draw, thick, isosceles triangle, minimum height = 3em,
        isosceles triangle apex angle=60]
\tikzstyle{input} = [coordinate]
\tikzstyle{output} = [coordinate]
\tikzstyle{pinstyle} = [pin edge={to-,thin,black}]

\begin{tikzpicture}[auto, node distance=2.5cm]

    \node [input, name=input] {};
    \node [block, right of=input] (hpf) 
    {$H_{HP}(z)$};
    \node [block, right of=hpf,
            node distance=3.25cm] (square_sig) {\Large{$c\cdot ( \text{} )^{2}$}};
    \node [block, right of=square_sig,
            node distance=2.5cm] (lpf) 
            {$H_{LP}(z)$};

    \draw [draw,->] (input) -- node {\large{$v_{i,t}$}} (hpf);
    \draw [->] (hpf) -- node[name=u1] {\large{$\Delta v_{i,t}$}} (square_sig);
    \draw [->] (square_sig) -- node[name=u2] {} (lpf);
    \node [output, right of=lpf] (output) {};
    \draw [->] (lpf) -- node [name=y] {\large{$y_{i,t}$}}(output);

\end{tikzpicture} }
\caption{Block diagram of instability detector using a transfer function representation of high and low pass filters.}
\label{fig:observer}
\end{figure}
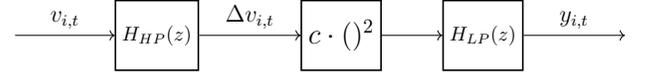

We propose the use of a simple filter to determine the ``energy" associated with voltage oscillations in the distribution grid.  The filter consists of the series  of a highpass filter, and an energy detector, consisting of a square-law, followed by a lowpass filter.  A discrete time block diagram of the process is shown in Fig. \ref{fig:observer}

where $H_{HP}$ and $H_{LP}$ are high-pass and low-pass filters respectively, realized using a bilinear transform equivalent of a first-order high/low-pass filter, and $c$ is a positive gain.  The high-pass filter removes DC content from $v_{i,t}$, yielding $ \Delta v_{i,t}$.  This signal is then squared to produce a DC term which is then extracted via low pass filtering.  The output signal, $y_i,t$ is a measure of the intensity of the instability.  The filter parameters should be chosen such that the filter does not attenuate oscillations due to inverter instabilities. 

\subsection{DER Cyber-Attack Mitigation as a RL problem}

The primary goal of the DRL controller is to mitigate instabilities introduced by DER smart inverter VV/VW controllers due to maliciously chosen set-points.  Let the graph $G = (\mathcal{N},\mathcal{L})$ represent the topology of the distribution feeder considered, where $\mathcal{N}$ is the set of nodes of the feeder (with 0 indexing the feeder head) and $\mathcal{L}$ is the set of lines.  For simplicity of presentation, we assume the presence of a VV/VW capable smart-inverter at every node in the system, so that the total number of inverters in the system is $|\mathcal{N}|$. We suppose the set $\mathcal{N}$ is partitioned into two sets, $\mathcal{H}$, and $\mathcal{U}$, where $\mathcal{H} \bigcup \mathcal{U} = \mathcal{N}$ which represent the "compromised" and '"uncompromised" inverters respectively. Furthermore we assume that $\mathcal{U} \neq \emptyset$, i.e. we have some controllable resources to mitigate the effects of the cyber-physical attack.
Given $\mathcal{U} \subsetneq  \mathcal{N}$ and the temporal dependency of load and solar irradiance, as mentioned in Remark 1, the model is a POMDP where we wish to determine the optimum stochastic policy, $\pi_{\theta}(a|o)$, parameterized by the neural network parameters $\theta$, modeling the probability distribution of action $a \in \mathcal{A}$ given the observation $o \in \mathcal{O}$.  

\begin{figure}[H]
    \centering
    \resizebox {17pc} {!} { \begin{tikzpicture}

\coordinate (xmax) at (5,0);
\coordinate (xmin) at (-5,0);
\draw[<->] (xmin) -- (xmax) node[anchor=west] {$V$};
    
\coordinate (ymax) at (0,3);
\coordinate (ymin) at (0,-3);
\draw[<->] (ymin) -- (ymax) node[anchor=south] {\% available VARs};
    
\filldraw[black]	
		(-2,0) circle (2pt) node[anchor=north] {$\eta_{2}$}
		(2,0) circle (2pt) node[anchor=south] {$\eta_{3}$}
        
        (-4,2) circle (2pt)
        (4,-2) circle (2pt);
        
\draw	
		(0,2) node[anchor=west] {100\%}
		(-4,0) node[anchor=north] {$\eta_{1}$}
		(0,-2) node[anchor=east] {-100\%}
		(4,0) node[anchor=south] {$\eta_{4}$};

\draw[dotted] (4,0) -- (4,-2);
\draw[dotted] (-4,0) -- (-4,2);

\draw[dotted] (0,-2) -- (4,-2);
\draw[dotted] (0,2) -- (-4,2);

\draw[thick][<->] (-5,2) -- (-4,2) -- (-2,0) -- (2,0) -- (4,-2) -- (5,-2);

\filldraw[red]	
		(-1.5,0) circle (2pt)
		(2.5,0) circle (2pt)
        (-3.5,2.0) circle (2pt)
        (4.5,-2.0) circle (2pt);

\draw[red, dashed] (-1.5, 0) -- (2.5,0);
\draw[red, dashed] (-1.5, 0) -- (-3.5,2);
\draw[red, dashed] (2.5, 0) -- (4.5,-2);
\draw[red, dashed] (-4.5, 2) -- (-3.5,2);

\draw(-3.8,3.2) node[anchor=north, align=center] {RL Agent Action\\$\Delta \eta$};
\draw[red, ->] (-4,2.2) -- (-3.5,2.2);





\end{tikzpicture}}
    \caption{Action example.}
    \label{fig:example_action}
\end{figure}
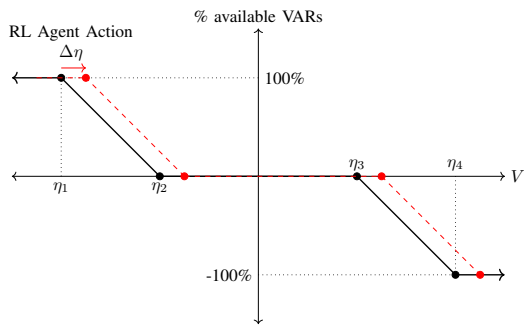

\noindent{\bf Training}: Rather than training multiple agents simultaneously, we adopt the following heuristics to aid convergence:
\begin{enumerate}
    \item For agent training, we define a single  agent whose input observation vector is the mean of the input observation vectors of all controllable inverters  $\in \mathcal{U}$ and whose action, $a_{t}$, is a deviation/offset, $\Delta \eta$, from default VV/VW control curves that apply across inverters.
    \item Once a single agent has been trained, this agent optimal policy is deployed locally on each individual inverter and only acts on local observations.
    \item Rather than optimize over arbitrarily shaped VV/VW curves ($f^q(\bar v)$ and $f^p(\bar v)$), we optimize over the deviation, i.e. $a=\Delta \eta$, from the default parameters defining the curves in Figs. \ref{fig:vvc} - \ref{fig:vwc}.  An example of this is shown in Fig. \ref{fig:example_action}.
    The translation is in range from -0.05 pu to 0.05 pu around an inverters default VV/VW curve, with the action space being discretized into $k$ bins.
    \item New parameterizations of VV/VW functions will be chosen so that measurement and power injection dynamics evolve on a faster timescale.  This choice will preserve the Markov property between actions taken by the RL controller.
\end{enumerate}
{\bf Observation}: The observation vectors $o_{i,t}, i\in \cal U$ at each RL agent (i.e. the input to the neural network that learns the optimum policy $\pi(a|o)$), consist of:
\begin{enumerate}
    \item $y_{i,t}$: the mean of the estimation of voltage oscillation energy at node $i$ since the last agent environment interaction.
    \item $y_{i,t}^{\text{max}}$: the maximum of $y_{i,t}$ over the previous $n$ environment interactions. This is a tunable parameter that stores information of the recent oscillation energy.
    \item $q^{\text{avail, nom}}_{i,t}$: the available reactive power capacity without active power curtailment.
    \item $a^\text{one-hot}_{i,t-1}$: one-hot encoding of the previous action taken by the agent.
\end{enumerate} 
%
{\bf Reward}: At a timestep $t$, the reward function, $R_{t}(a_t,o_t)$ is:
\begin{align}
R_{t} = &-\Bigg(\frac{1}{|\mathcal{U}|}\sum_{i=1}^{|\mathcal{U}|}\sigma_y y_{i,t} + \sigma_{a} \mathbbm{1}_{a_t\neq a_{t-1}} + \sigma_{0} \lVert a_t \rVert_{2} \nonumber \\
&+ \frac{1}{|\mathcal{U}|}\sum_{i=1}^{|\mathcal{U}|}\sigma_{p} \left(1 - \frac{p_{i,t}}{p_{i,t}^{\text{max}}}\right)^2 \Bigg). \label{eq:rt}
\end{align}
The first component seeks to minimize the voltage oscillation $y$; the second one penalizes configuration changes on inverters; the third component encourages the agent to use the default inverter configurations in the absence of voltage oscillations and the final component penalizes any active power curtailment.
%
%



\subsection{The PyCIGAR DRL Environment}
Any learning method requires sufficient training over a variety of scenarios. As is done in other application of deep learning in the context of critical infrastructure systems, such training can be performed through realistic Monte Carlo simulations that cover a variety of operating conditions and cyber-physical attacks. We named PyCIGAR\footnote{The name stands for Python based Cybersecurity via Inverter-Grid Automatic Reconfiguration.} the modular software architecture we designed to train the DRL agent described in the previous section.  
%
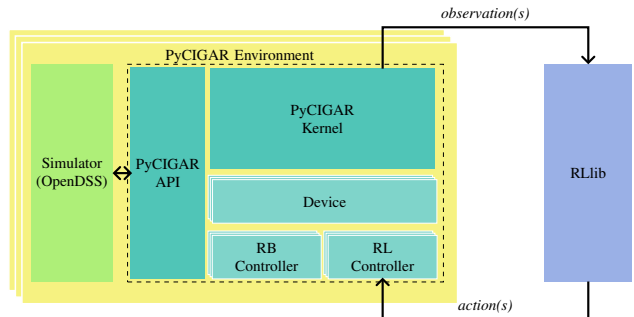
\begin{figure}[H]
    \centering
    \resizebox {20pc} {!} { \begin{tikzpicture}
\definecolor{c5}{RGB}{246,242,129}
\definecolor{c5_}{RGB}{244,240,118}
\definecolor{c4}{RGB}{174,239,120}
\definecolor{c3}{RGB}{80,197,183}
\definecolor{c3_}{RGB}{127,212,202}
\definecolor{c2}{RGB}{154,176,230} 
\definecolor{c1}{RGB}{83,58,113}

\tikzstyle{pycigar_node}=[rectangle, align=center, minimum width=1cm, minimum height=4.85cm]
\tikzstyle{simulator_node}=[rectangle, align=center, minimum width=1cm, minimum height=5cm]
\tikzstyle{rllib_node}=[rectangle, align=center, minimum width=2cm, minimum height=5cm]
\tikzstyle{env_node}=[rectangle, align=center, 
                      minimum width=10cm, minimum height=6cm, 
                      label={[anchor=north, inner sep=5pt] north:PyCIGAR Environment}, 
                      draw=white, line width=1pt,]
\tikzstyle{dashed_node}=[rectangle, dashed, align=center, 
                        minimum width=7.25cm, minimum height=5cm, 
                      draw=black, line width=0.5pt,]

\tikzstyle{device_node}=[rectangle, align=center, minimum width=5.15cm, minimum height=1cm,
                        draw=white, line width=0.5pt,]

\tikzstyle{control_node}=[rectangle, align=center, minimum width=2.5cm, minimum height=1cm,
                        draw=white, line width=0.5pt,]
                        
\tikzstyle{kernel_node}=[rectangle, align=center, minimum width=5.15cm, minimum height=2.3cm]
                        
\node[env_node] (env3) at (-3.30,0.30) [fill=c5] {};
\node[env_node] (env2) at (-3.15,0.15) [fill=c5] {};
\node[env_node] (env) at (-3,0) [fill=c5] {};

\node[dashed_node] (dashed) at (-1.95,0) {}; 

\node[simulator_node] (simulator) at (-6.85,0) [fill=c4] {Simulator\\(OpenDSS)};
\node[pycigar_node] (pycigar) at (-4.65,0) [fill=c3]{PyCIGAR \\API};
\node[rllib_node] (rllib) at (5,0) [fill=c2] {RLlib};

\node[device_node] (device3) at (-1.15,-1.05 + 0.5) [fill=c3_] {Device};
\node[device_node] (device2) at (-1.1,-1.1   + 0.5) [fill=c3_] {Device};
\node[device_node] (device) at (-1.05,-1.15  + 0.5) [fill=c3_] {Device};

\node[control_node] (control3) at (-1.15 - 1.33,-1.05 - 0.78) [fill=c3_] {RB \\ Controller};
\node[control_node] (control2) at (-1.1  - 1.33,-1.1   - 0.78) [fill=c3_] {RB \\ Controller};
\node[control_node] (control) at (-1.05  - 1.33,-1.15  - 0.78) [fill=c3_] {RB \\ Controller};

\node[control_node] (control3) at (-1.15 + 1.33,-1.05 - 0.78) [fill=c3_]  {RL \\ Controller};
\node[control_node] (control2) at (-1.1  + 1.33,-1.1   - 0.78) [fill=c3_] {RL \\ Controller};
\node[control_node] (control) at (-1.05  + 1.33,-1.15  - 0.78) [fill=c3_] {RL \\ Controller};

\node[kernel_node] (kernel) at (-1.1,1 + 0.25) [fill=c3] {PyCIGAR \\ Kernel};

\draw[line width=0.5mm,black,angle 90-] (-1.05  + 1.33,-1.15  - 1.25) -- (-1.05  + 1.33,-1.15  - 2.2) -- node[pos=0.5,sloped,above] {\textit{action(s)}} (5,-1.15  - 2.2) -- (5,-1.15  - 1.34); 
\draw[line width=0.5mm,black,-angle 90] (-1.05 + 1.33,-1.15 - 1.25 + 4.8) -- (-1.05 + 1.33,-1.15 - 2.2 + 6.7) -- node[pos=0.5,sloped,above] {\textit{observation(s)}} (5,-1.15 - 2.2 + 6.7) -- (5,2.5);
\draw[line width=0.5mm,black=c1,angle 90-angle 90] (0 -5.92,0) -- (0.45 - 5.92,0);
\end{tikzpicture}}
    \caption{PyCIGAR Architecture.}
    \label{fig:pycigar_architecture}
\end{figure}
PyCIGAR is a Python library for distributed reinforcement learning for electric power distribution grids on quasi-static time scales.  The library provides a link between power system simulators and a reinforcement learning library - RLlib \cite{liang2017rllib}. PyCIGAR is a unified API that can interface different power system simulators (e.g. OpenDSS), while on the RL side PyCIGAR uses RLlib in order to deploy large scale experiments on a server, machine cluster or on a cloud. 

A diagram of the PyCIGAR architecture is shown in Fig. \ref{fig:pycigar_architecture}.  In addition to RL-based controllers, PyCIGAR also includes rule-based (RB) control devices (e.g. tap-changing transformers) and can easily be extended to support the integration of other more complicated DER (e.g. electric vehicle charging and battery storage systems). PyCIGAR provides a foundation for the rapid development of learning-based control algorithms for heterogeneous classes of DER in electric power distribution grids.

\section{Results}
\label{sec:results}



We conduct experiments on the IEEE 37-bus feeder with all load buses having an active power generation of 50\% of the nominal load with an additional 10\% inverter over-sizing for reactive power headroom. The agent training environment consists of 700 one-second timesteps per simulation. At the end of each experiment, the training environment is reset with randomized load and solar generation profiles and percentage of compromised inverters.  This diversity creates a rich environment that exposes the RL agent to attacks that could occur anytime throughout the day under a variety of loading, solar conditions and proportions of compromised inverters. For each case, all inverters start with their default VV/VW settings and at a particular time in the simulation the attacker gains controls of $15\%$ to $50\%$ of the installed inverter capacity at each node to create a voltage instability. It does so by translating the VV/VW curves and steepening the slopes to induce an oscillation. This attack vector represents a subset of possible attack vectors. The agent is allowed to reconfigure the VV /VW curves of non-compromised DER to mitigate oscillations that result from the cyber-attack. We consider two types of action, 1) translating the entire VV/VW piecewise functions from its default configuration (offset action) and 2) adjusting the slope of the piece wise function in the region  $\in (\eta_{1},  \eta_{2}]$ and $\in (\eta_{3},  \eta_{4}]$ (slope action). Within the simulation, the agent receives observations and updates inverters' functions $f^q_{i}$ and $f^p_{i}$ every 35 seconds.  The training is conducted on an Intel® Xeon® E5-2623 v3 processor, 64GB RAM server and takes 1 hour of training time to converge. 

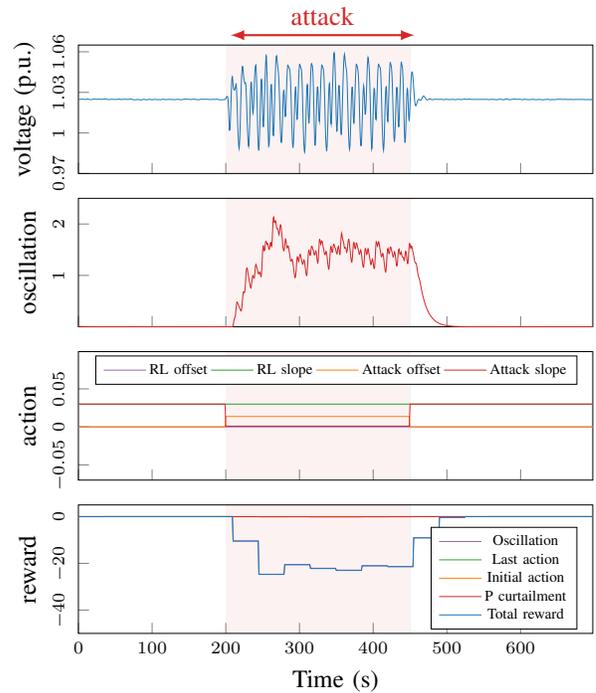
\begin{figure}[ht]
\centering
\def\evalhist{data/eval_hist_no_act_new_new_filter_solar_randomization_8f_new.csv}
\def\hackstart{200}
\def\hackend{450}
\begin{tikzpicture}[scale=\scale]


\definecolor{tableau_blue}{RGB}{31,119,180}
\definecolor{tableau_orange}{RGB}{255,127,14}
\definecolor{tableau_red}{RGB}{214,39,40}
\definecolor{tableau_grey}{RGB}{127,127,127}
\definecolor{tableau_green}{RGB}{44,160,44}
\definecolor{tableau_purple}{RGB}{148,103,189}
\definecolor{tableau_cyan}{RGB}{23,190,207}

\newcommand{\hackstartline}{
\draw [tableau_red, very thick, dashed] 
(axis cs:\hackstart,\pgfkeysvalueof{/pgfplots/ymin}) -- 
(axis cs:\hackstart,\pgfkeysvalueof{/pgfplots/ymax});}

\newcommand{\hackendline}{
\draw [tableau_red, very thick, dashed] 
(axis cs:\hackend,\pgfkeysvalueof{/pgfplots/ymin}) -- 
(axis cs:\hackend,\pgfkeysvalueof{/pgfplots/ymax});}

\newcommand{\hacklines}{\hackstartline\hackendline}

\newcommand{\hackshade}{
\draw [fill=tableau_red, opacity=.06] 
(axis cs:\hackstart,\pgfkeysvalueof{/pgfplots/ymin}) -- 
(axis cs:\hackend,\pgfkeysvalueof{/pgfplots/ymin}) -- 
(axis cs:\hackend,\pgfkeysvalueof{/pgfplots/ymax}) -- 
(axis cs:\hackstart,\pgfkeysvalueof{/pgfplots/ymax});
}

\begin{groupplot}[group style={
        group name=my plots,
        group size=1 by 4,
        xlabels at=edge bottom,
        xticklabels at=edge bottom,
        vertical sep=10pt,
    },
    xlabel={Time (s)},
    legend style={nodes={scale=0.6, transform shape}},
    width=\linewidth,
    height=3.4cm,
    enlarge x limits=false,
    tick pos=left,
    xticklabel style={font=\scriptsize},
    yticklabel style={font=\scriptsize,rotate=90,anchor=base,yshift=0.2cm,scaled y ticks=false, /pgf/number format/fixed},
    minor y tick num=0,
]
\nextgroupplot[ylabel=voltage (p.u.),legend pos=north east, ymax=1.065, ymin=0.97,ytick={0.97, 1, 1.03, 1.06}]\hackshade
\addplot [smooth, tableau_blue, no markers] table [x=, y=v, col sep=comma] {\evalhist};
\coordinate (hackstart) at (axis cs:\hackstart*\scale,\pgfkeysvalueof{/pgfplots/ymax});
\coordinate (hackend) at (axis cs:\hackend*\scale,\pgfkeysvalueof{/pgfplots/ymax});


\nextgroupplot[ylabel=oscillation,legend pos=north east, ymin=0, ymax=2.5, ytick={1.0, 2.0}]\hackshade
\addplot [smooth, tableau_red, no markers] table [x=, y=y, col sep=comma] {\evalhist};


\nextgroupplot[ylabel=action, legend pos=north east, legend columns=4,try min ticks=1,ymin=-0.07 , ymax=0.099]\hackshade
\addplot [tableau_purple, no markers] table [x=, y=translation, col sep=comma] {\evalhist};
\addplot [tableau_green, no markers] table [x=, y=slope, col sep=comma] {\evalhist};
\addplot [tableau_orange, no markers] table [x=, y=translation_adv, col sep=comma] {\evalhist};
\addplot [tableau_red, no markers] table [x=, y=slope_adv, col sep=comma] {\evalhist};


\addlegendentry{RL offset}
\addlegendentry{RL slope}
\addlegendentry{Attack offset}
\addlegendentry{Attack slope}

\nextgroupplot[ylabel=reward, legend pos=south east, legend columns=1, ymax=5, ymin=-50]\hackshade
\addplot [tableau_purple, no markers] table [x=, y=component_y, col sep=comma] {\evalhist};
\addplot [tableau_green, no markers] table [x=, y=component_oa, col sep=comma] {\evalhist};
\addplot [tableau_orange, no markers] table [x=, y=component_init, col sep=comma] {\evalhist};
\addplot [tableau_red, no markers] table [x=, y=component_pset_pmax, col sep=comma] {\evalhist};
\addplot [tableau_blue, no markers] table [x=, y=total_reward, col sep=comma] {\evalhist};

\addlegendentry{Oscillation}
\addlegendentry{Last action}
\addlegendentry{Initial action}
\addlegendentry{P curtailment}
\addlegendentry{Total reward}

\end{groupplot}

\draw[latex-latex, thick, tableau_red, yshift=0.5cm] (hackstart) to node [auto] {attack} (hackend);

\end{tikzpicture}

\caption{Result of an evaluation episode at 45\% attack without agent defense}
\label{fig:hack45noact}
\end{figure}

Fig. \ref{fig:hack45noact} shows the baseline case caused by a 45\% percentage attack around noon with no action taken to mitigate the result of the attack.  The attack creates oscillations in system voltages that are detected by the oscillation detector (see Section \ref{subsec:detect}).  The malicious re-dispatch of settings are shown in the action subplot and the components of the reward function, \eqref{eq:rt}, are shown at the bottom. In the absence of an control the reward is solely composed of the penalty for the oscillation. 

\begin{figure}[ht]
\centering

\def\evalhist{data/eval_hist_11000_0.45_new_new_filter_solar_randomization_8f_new.csv}
\def\hackstart{200}
\def\hackend{450}
\begin{tikzpicture}[scale=\scale]


\definecolor{tableau_blue}{RGB}{31,119,180}
\definecolor{tableau_orange}{RGB}{255,127,14}
\definecolor{tableau_red}{RGB}{214,39,40}
\definecolor{tableau_grey}{RGB}{127,127,127}
\definecolor{tableau_green}{RGB}{44,160,44}
\definecolor{tableau_purple}{RGB}{148,103,189}
\definecolor{tableau_cyan}{RGB}{23,190,207}

\newcommand{\hackstartline}{
\draw [tableau_red, very thick, dashed] 
(axis cs:\hackstart,\pgfkeysvalueof{/pgfplots/ymin}) -- 
(axis cs:\hackstart,\pgfkeysvalueof{/pgfplots/ymax});}

\newcommand{\hackendline}{
\draw [tableau_red, very thick, dashed] 
(axis cs:\hackend,\pgfkeysvalueof{/pgfplots/ymin}) -- 
(axis cs:\hackend,\pgfkeysvalueof{/pgfplots/ymax});}

\newcommand{\hacklines}{\hackstartline\hackendline}

\newcommand{\hackshade}{
\draw [fill=tableau_red, opacity=.06] 
(axis cs:\hackstart,\pgfkeysvalueof{/pgfplots/ymin}) -- 
(axis cs:\hackend,\pgfkeysvalueof{/pgfplots/ymin}) -- 
(axis cs:\hackend,\pgfkeysvalueof{/pgfplots/ymax}) -- 
(axis cs:\hackstart,\pgfkeysvalueof{/pgfplots/ymax});
}

\begin{groupplot}[group style={
        group name=my plots,
        group size=1 by 4,
        xlabels at=edge bottom,
        xticklabels at=edge bottom,
        vertical sep=10pt,
    },
    xlabel={Time (s)},
    legend style={nodes={scale=0.6, transform shape}},
    width=\linewidth,
    height=3.4cm,
    enlarge x limits=false,
    tick pos=left,
    xticklabel style={font=\scriptsize},
    yticklabel style={font=\scriptsize,rotate=90,anchor=base,yshift=0.2cm,scaled y ticks=false, /pgf/number format/fixed},
    minor y tick num=0,
]
\nextgroupplot[ylabel=voltage (p.u.),legend pos=north east, ymax=1.065, ymin=0.97,ytick={0.97, 1, 1.03, 1.06}]\hackshade
\addplot [smooth, tableau_blue, no markers] table [x=, y=v, col sep=comma] {\evalhist};
\coordinate (hackstart) at (axis cs:\hackstart*\scale,\pgfkeysvalueof{/pgfplots/ymax});
\coordinate (hackend) at (axis cs:\hackend*\scale,\pgfkeysvalueof{/pgfplots/ymax});


\nextgroupplot[ylabel=oscillation,legend pos=north east, ymin=0, ymax=2.5, ytick={1.0, 2.0}]\hackshade
\addplot [smooth, tableau_red, no markers] table [x=, y=y, col sep=comma] {\evalhist};


\nextgroupplot[ylabel=action, legend pos=north east, legend columns=4,try min ticks=1,ymin=-0.07 , ymax=0.099]\hackshade
\addplot [tableau_purple, no markers] table [x=, y=translation, col sep=comma] {\evalhist};
\addplot [tableau_green, no markers] table [x=, y=slope, col sep=comma] {\evalhist};
\addplot [tableau_orange, no markers] table [x=, y=translation_adv, col sep=comma] {\evalhist};
\addplot [tableau_red, no markers] table [x=, y=slope_adv, col sep=comma] {\evalhist};


\addlegendentry{RL offset}
\addlegendentry{RL slope}
\addlegendentry{Attack offset}
\addlegendentry{Attack slope}

\nextgroupplot[ylabel=reward, legend pos=south east, legend columns=1, ymax=5, ymin=-50]\hackshade
\addplot [tableau_purple, no markers] table [x=, y=component_y, col sep=comma] {\evalhist};
\addplot [tableau_green, no markers] table [x=, y=component_oa, col sep=comma] {\evalhist};
\addplot [tableau_orange, no markers] table [x=, y=component_init, col sep=comma] {\evalhist};
\addplot [tableau_red, no markers] table [x=, y=component_pset_pmax, col sep=comma] {\evalhist};
\addplot [tableau_blue, no markers] table [x=, y=total_reward, col sep=comma] {\evalhist};

\addlegendentry{Oscillation}
\addlegendentry{Last action}
\addlegendentry{Initial action}
\addlegendentry{P curtailment}
\addlegendentry{Total reward}

\end{groupplot}

\draw[latex-latex, thick, tableau_red, yshift=0.5cm] (hackstart) to node [auto] {attack} (hackend);

\end{tikzpicture}

\caption{Result of an evaluation episode at 45\% attack around noon}
\label{fig:hack45}
\end{figure}
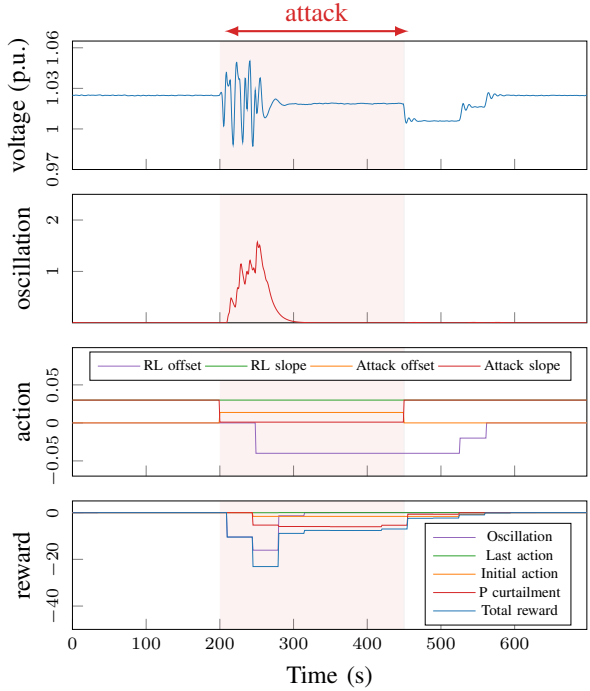

Fig. \ref{fig:hack20} - \ref{fig:hack45}  show the behavior from the trained RL agent at a random node in the network in mitigating instabilities from compromised DER at two different times of day and different percentage of compromised DER. At simulation time $t=200\,\text{s}$ the attack is introduced in a portion of DER. This can be seen in the action subplot, which shows the breakpoints of the piecewise linear curves of \ref{fig:vvc} - \ref{fig:vwc} being suddenly moved to a new configuration. This triggers an oscillation in grid voltages. The output of the \emph{oscillation observer} is both fed into the RL agent as an observation and included as a negative penalty in the reward function. The agent, therefore, should control non-compromised assets to minimize the oscillation.  This is what occurs, as the agent changes the breakpoints of non-compromised units just after $t=250\,\text{s}$ by translating the default VV/VW curves. This action almost immediately stops the oscillation in the system voltages, resulting in a defeat of the original cyber-attack. Fig. \ref{fig:hack20} features an attack in the morning, around 9am, where there is significant excess capacity for reactive power compensation available for the agent to mitigate the cyber-attack. The agent, therefore, does not need to curtail active power generation to successfully mitigate the attack. This, however, is not the case in Fig. \ref{fig:hack45} where the attack occurs around midday and the agent is forced to curtail active power in order to have enough controllability to defeat the attack. Across numerous training configurations it was observed that RL offset was the preferred action of the agent. 

Also worth highlighting is the behavior of the agent after the compromised DER have been identified and returned to their original settings, at $t=450\,\text{s}$.  Shortly after, we observe the agent also returning to its original configuration.  In demonstrating this behavior, it can be seen that the RL agent will take steps to ameliorate the effects of the cyber attack, but will return a state of inactivity once the threat of the attack has passed.


\begin{figure}[ht]
\centering

\def\evalhist{data/eval_hist_100_0.2_new_new_filter_solar_randomization_8f_new.csv}
\def\hackstart{200}
\def\hackend{450}
\begin{tikzpicture}[scale=\scale]


\definecolor{tableau_blue}{RGB}{31,119,180}
\definecolor{tableau_orange}{RGB}{255,127,14}
\definecolor{tableau_red}{RGB}{214,39,40}
\definecolor{tableau_grey}{RGB}{127,127,127}
\definecolor{tableau_green}{RGB}{44,160,44}
\definecolor{tableau_purple}{RGB}{148,103,189}
\definecolor{tableau_cyan}{RGB}{23,190,207}

\newcommand{\hackstartline}{
\draw [tableau_red, very thick, dashed] 
(axis cs:\hackstart,\pgfkeysvalueof{/pgfplots/ymin}) -- 
(axis cs:\hackstart,\pgfkeysvalueof{/pgfplots/ymax});}

\newcommand{\hackendline}{
\draw [tableau_red, very thick, dashed] 
(axis cs:\hackend,\pgfkeysvalueof{/pgfplots/ymin}) -- 
(axis cs:\hackend,\pgfkeysvalueof{/pgfplots/ymax});}

\newcommand{\hacklines}{\hackstartline\hackendline}

\newcommand{\hackshade}{
\draw [fill=tableau_red, opacity=.06] 
(axis cs:\hackstart,\pgfkeysvalueof{/pgfplots/ymin}) -- 
(axis cs:\hackend,\pgfkeysvalueof{/pgfplots/ymin}) -- 
(axis cs:\hackend,\pgfkeysvalueof{/pgfplots/ymax}) -- 
(axis cs:\hackstart,\pgfkeysvalueof{/pgfplots/ymax});
}

\begin{groupplot}[group style={
        group name=my plots,
        group size=1 by 4,
        xlabels at=edge bottom,
        xticklabels at=edge bottom,
        vertical sep=10pt,
    },
    xlabel={Time (s)},
    legend style={nodes={scale=0.6, transform shape}},
    width=\linewidth,
    height=3.4cm,
    enlarge x limits=false,
    tick pos=left,
    xticklabel style={font=\scriptsize},
    yticklabel style={font=\scriptsize,rotate=90,anchor=base,yshift=0.2cm,scaled y ticks=false, /pgf/number format/fixed},
    minor y tick num=0,
]
\nextgroupplot[ylabel=voltage (p.u.),legend pos=north east, ymax=1.065, ymin=0.97,ytick={0.97, 1, 1.03, 1.06}]\hackshade
\addplot [smooth, tableau_blue, no markers] table [x=, y=v, col sep=comma] {\evalhist};
\coordinate (hackstart) at (axis cs:\hackstart*\scale,\pgfkeysvalueof{/pgfplots/ymax});
\coordinate (hackend) at (axis cs:\hackend*\scale,\pgfkeysvalueof{/pgfplots/ymax});


\nextgroupplot[ylabel=oscillation,legend pos=north east, ymin=0, ymax=2.5, ytick={1.0, 2.0}]\hackshade
\addplot [smooth, tableau_red, no markers] table [x=, y=y, col sep=comma] {\evalhist};


\nextgroupplot[ylabel=action, legend pos=north east, legend columns=4,try min ticks=1,ymin=-0.07 , ymax=0.099]\hackshade
\addplot [tableau_purple, no markers] table [x=, y=translation, col sep=comma] {\evalhist};
\addplot [tableau_green, no markers] table [x=, y=slope, col sep=comma] {\evalhist};
\addplot [tableau_orange, no markers] table [x=, y=translation_adv, col sep=comma] {\evalhist};
\addplot [tableau_red, no markers] table [x=, y=slope_adv, col sep=comma] {\evalhist};


\addlegendentry{RL offset}
\addlegendentry{RL slope}
\addlegendentry{Attack offset}
\addlegendentry{Attack slope}

\nextgroupplot[ylabel=reward, legend pos=south east, legend columns=1, ymax=5, ymin=-50]\hackshade
\addplot [tableau_purple, no markers] table [x=, y=component_y, col sep=comma] {\evalhist};
\addplot [tableau_green, no markers] table [x=, y=component_oa, col sep=comma] {\evalhist};
\addplot [tableau_orange, no markers] table [x=, y=component_init, col sep=comma] {\evalhist};
\addplot [tableau_red, no markers] table [x=, y=component_pset_pmax, col sep=comma] {\evalhist};
\addplot [tableau_blue, no markers] table [x=, y=total_reward, col sep=comma] {\evalhist};

\addlegendentry{Oscillation}
\addlegendentry{Last action}
\addlegendentry{Initial action}
\addlegendentry{P curtailment}
\addlegendentry{Total reward}

\end{groupplot}

\draw[latex-latex, thick, tableau_red, yshift=0.5cm] (hackstart) to node [auto] {attack} (hackend);

\end{tikzpicture}

\caption{Result of an evaluation episode at 20\% attack around 9AM}
\label{fig:hack20}
\end{figure}
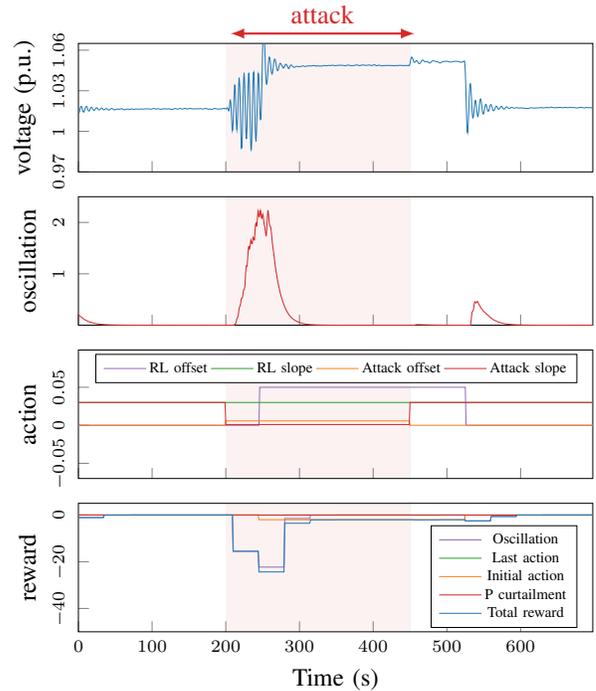

\section{Conclusions}
\label{sec:conclusions}
This paper has proposed a reinforcement learning approach for mitigating the oscillation due to unstable smart inverter settings by training an agent to translate the VV/VW curves. The resultant policy successfully mitigated adversary induced voltage oscillatory behavior for the cases considered.

Future work will investigate the value of this approach for larger networks and the sensitivity of the trained agents to specific network topologies/configuration. Additionally, we will explore different neural network architectures, including Recurrent Neural Network (RNN) and Long Short-Term Memory (LSTM), which have proven to be the state of the art in solar and load forecasting and may improve the performance of the agent. Additional types of attacks will also be considered, including, but not limited to, voltage imbalance attacks. An adversary may seek to exploit DER interaction with utility voltage regulation systems to create system voltage imbalances, leading to device trips and possible system collapse.

\bibliographystyle{IEEEtran}
\bibliography{references}

\appendix

\renewcommand{\arraystretch}{1.25}
\begin{table}[h!]
\centering
\begin{tabular}{|ll|}
\hline
\textbf{Hyperparameter} & \textbf{Value} \\
\hline

$\alpha$ (learning rate)                       & $1\times 10^{-3}$                    \\
$\gamma$ (reward discount factor)                            & 0.5                     \\
$\lambda$ (GAE parameter)                           & 0.95                    \\
$\epsilon$ (PPO clip param)                            & 0.1 \\
batch size                          & 420                     \\ 
activation function                 & tanh                    \\
network hidden layers               & dense (64, 64, 32) \\ 
$\sigma_y$ (oscillation penalty)             & 15                        \\
$\sigma_a$ (action penalty)                  & 0.05                        \\
$\sigma_0$ (penalty for deviation from default VV/VW curve) & 18                       \\
$\sigma_p$ (penalty for curtailing active power)          & 80                        \\ 
action range               &$-0.05$\,pu to 0.05\,pu \\
$k$ (action range discretization)               &0.01 p.u. \\

\hline

\end{tabular}
\vspace{0.1cm}
\caption{Hyperparameters of the network, training and reward }
\label{hyperparams}
\end{table}

\end{document}